# Detecting domain wall trapping and motion at a constriction in narrow ferromagnetic wires using perpendicular-current giant magnetoresistance

A. J. Zambano and W. P. Pratt, Jr.

*Department of Physics and Astronomy, Michigan State University, East Lansing, MI 48824-2320*

We present a versatile method for detecting the presence and motion of a trapped domain wall in a narrow ferromagnetic layer using current-perpendicular-to-plane (CPP) giant magnetoresistance (MR). The CPP-MR response to small motions of the trapped domain wall is enhanced because the CPP current is restricted to the region of wall trapping. We use a Permalloy/Cu/Permalloy spin valve in the shape of a long, ~500-nm-wide wire with a constriction (notch) near its middle that acts as a trapping site for a head-to-head domain wall. Two different notch shapes were studied, mostly at 4.2 K but also at 295K.

Control and detection of geometrically confined domain walls are of intense interest both for physics and device applications. Ono *et al.*[1] used a $Ni_{81}Fe_{19}(20nm)/Cu(10nm)/Ni_{81}Fe_{19}(5nm)$ spin valve to trap a domain wall in a specific place and detect the wall's presence using current-in-plane giant magnetoresistance (MR). The spin valve was in the shape of a long, 0.5-μm-wide 'wire' that had voltage contacts 20 μm apart, between which a constriction (0.35-μm wide) was placed. This 'notch' acted as a trapping site for a domain wall (see Fig. 1(a)). The shape anisotropy of the wire constrained the magnetizations (M) of both NiFe layers to be parallel (P) to wire axis. After application of a strong external magnetic field (H) along the +x direction, H was then applied along −x, causing a 'head-to-head' domain wall to propagate in the thin NiFe layer until the wall became trapped in the notch as M reversed on one side of the notch. M of the thick NiFe layer remained fixed along +x. Since GMR depends on the relative orientation of M in the two NiFe layers, the presence of a trapped wall could be detected because the resistance is between the limits of both M's being parallel or antiparallel (AP). More recently, the emphasis has shifted to using high in-plane current densities to manipulate the wall motion in such narrow structures.[3-5] Kläui et al.[4] used a narrow single-layer ferromagnetic ring with notches where head-to-head walls could be manipulated by the orientation of an in-plane magnetic field and by high current densities circulating around the ring. Anisotropic magnetoresistance (AMR) detected the presence of walls, giving increased sensitivity to details of the wall trapping in the notches because the contribution to the total resistance was largest there.

We present a scheme for detecting the presence and motion of a trapped domain wall that provides enhanced sensitivity to small wall motions. We use current-perpendicular-to-plane (CPP) giant MR and confine the CPP current flow to the region of wall trapping. This method has the added flexibility of allowing the CPP current region to be chosen independently of one of the dimensions of the trapping region.

We have adapted the Ono *et al.*[1] method for trapping domain walls to CPP-MR measurements, as shown in Fig. 1(a, b) with top and side views of the patterned multilayer structure. The multilayer sequence is $Nb(100)/Cu(10)/Py(20)/Cu(10)/Py(5)/Cu(5)/Au(15)$, where $Py = Ni_{84}Fe_{16}$ and the thickness are in nm. We shape the Py/Cu/Py/Au part of the multilayer into a long 500-nm-wide long 'wire' with a constriction near its middle. The reservoir acts as a source for injecting domain walls into the wire.[2] H is applied parallel to the axis of the wire. The trapping at the notch of a head-to-head wall in the thin Py layer is depicted. The CPP current is confined to the notch region by a 900-nm-wide top contact, made of Au(200nm) or Nb(150nm). At 4.2 K, the latter provides an almost-uniform CPP current density because the bottom and top Nb layers are superconducting.[6] Fig. 1(c)-(i) & -(ii), show scanning electron microscope images of representative Al ion-etching masks for two notch configurations: 'U' and 'V', respectively. The length of the U-notch is comparable to the width of the top contact, while the length of the V-notch is smaller.

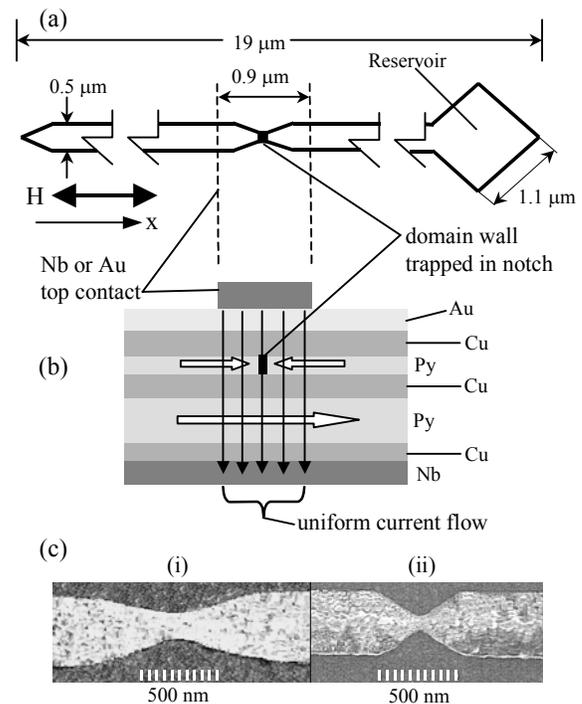

Fig. 1. (a) & (b): cartoon showing top and side views of a multilayer 'wire' with a head-to-head domain wall trapped in the thin Py layer at the notch. (c)-(i) &-(ii): SEM images of Al ion-milling masks for etching 'U' and 'V' shaped notches, respectively.

First the Nb/···/Au multilayer was sputter deposited onto a Si substrate. Next e-beam lithography, evaporation of a 300-nm-thick Al layer and lift off were used to define Al masks for subsequent ion milling to form the 'wire' and its notch. After ion milling through the thick Py layer, a planar layer of SiO was evaporated to insulate the top and bottom contacts. The Al mask was removed by a chemical etch. Finally, e-beam lithography, sputter deposition and liftoff were used to produce the Nb or Au top contact. Four-terminal resistance measurements, each of 1 to 3 s duration, were made using a lock-in amplifier operating at 8 kHz with current excitations $\leq 0.2$ mA rms.

Fig. 2 shows the CPP resistance R for a V-shaped notch. Near + 500 Oe, the cartoon shows that M of the thin and thick Py layers are in the P state and pointing to the right (filled arrows along +x). This is the low-resistance ($R_P$) state. The field is then lowered until a head-to-head domain wall is injected into the thin Py layer of the wire from the reservoir at the right (see Fig. 1(a)) at −10 Oe and becomes trapped at the notch. Such low injection fields agree with earlier work by Ono et al.[1] For the thin Py layer, M on the right side of the wall is reversed, as shown (open arrow along −x). As H is lowered further, the CPP-MR detects an average motion of the trapped wall to the left. At H = −100 Oe, the wall leaves the notch, and two Ms are now in the AP, high-R state ($R_{AP}$). Further reduction of H causes a head-to-head wall to be injected and trapped in the thick Py layer at H = −230 Oe. Finally, by −400 Oe, the wall in the thick layer has left the notch, and the Ms are now in the P state. Starting at −500 Oe and increasing the field gives very similar trapping behavior in R for the thin Py layer, but with 'tail-to-tail' walls.

The response of the trapped wall to H in the thin Py layer produces a linear variation of R that has a maximum value $\delta R \sim 10\% \times (R_{AP} - R_P)$. Micromagnetic simulations have been done for a head-to-head wall of transverse-magnetization ($\perp$ x) that is trapped in a trapezoidal notch cut into one side of a narrow wire in a very thin Py layer.[7] Such simulations indicate that the width (along x) of this transverse-M region for the head-to-head wall ranges approximately from the width ($\perp$ x) of the notch when the wall is trapped in the notch to the width of wire when it's outside the notch.[7,8] One can use $\delta R$ to put a constraint on the relationship between the width of this transverse-M region and the distance the trapped wall moves along x as the field changes. We adopt a very crude model where the head-to-head wall (of fixed width a) has uniform M transverse to x and elsewhere M is either parallel or antiparallel to x. If we let a = 100nm ($\sim$ the width of the notch $\perp$ to x), then the $\delta R$ corresponds to a $\sim$ 150 nm displacement of the wall. If we choose an upper bound of a = 500 nm (comparable to the width of the wire), then the displacement is $\sim$ 70 nm. A complete micromagnetics calculation of our V notch, combined with the uniform CPP current and constrained by $\delta R$, should provide a test of the micromagnetics, as was done using AMR.[4]

In Fig. 2, for $|H| \leq 100$ Oe and no trapping, the MR readily detects a subtle readjustment of the Ms away from the P state for the two Py layers. This is likely due to residual dipolar coupling between these two layers in the notch region. To confirm that this behavior was not associated with the wire-shaped Py layers outside the notch, we placed an extra Nb top contact far to the left of a U-shaped notch in a different sample (see Fig. 1(a)). Fig. 3 shows at 4.2 K that, for -70 Oe $\leq$ H $\leq$ +300 Oe, $\Delta R = (R - R_P) = 0$, indicating a lack of residual dipole coupling in the notch-free part of the wire. At H = -70 Oe, the wall leaves the U-notch and begins entering the region of this extra Nb top contact. A separate CPP-MR measurement at the U-notch of this sample established that the wall did become trapped in this notch at H $\approx$ -10 Oe and left the notch at H $\approx$ -70 Oe. For −110 Oe $\leq$ H $\leq$ -70 Oe, Fig. 3 exhibits some evidence of trapping at the extra top contact before the AP state is attained, presumably due to stress introduced by the top Nb contact during cooling of the sample. At 295K, the expected lower stress and higher thermal activation allow a more clean transition to the AP state with less trapping of the wall. These data also show that CPP detection of wall motion can be done even at 295K.

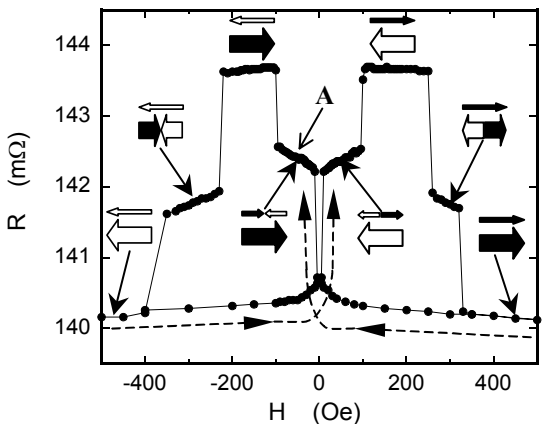

Fig. 2. CPP resistance R vs magnetic field H at 4.2 K for a V-shaped notch with a Nb top contact. Thick and thin arrows represent magnetizations of thick and thin Py layers, respectively. Dashed arrows indicate change in H. A refers to data displayed in Fig. 5.

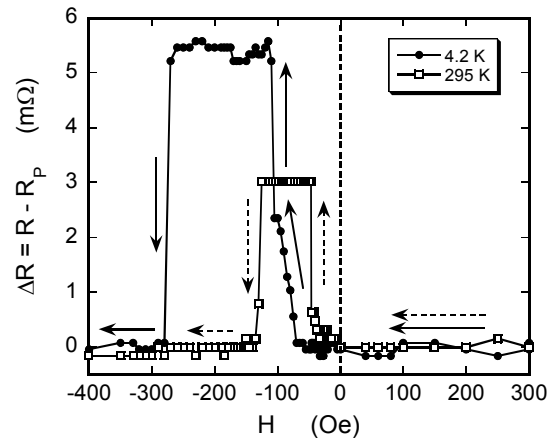

Fig. 3. $\Delta R$ (= R – $R_P$) vs H for a sample with an extra Nb top contact placed to the left of a U-shaped notch. Solid and dashed arrows show direction of field sweep at 4.2 K and 295 K, respectively.

In Fig. 4 we present $\Delta R$ vs H at 4.2 K for a U-shaped notch with a Au top contact. For the sample in Fig. 3, $R_P \approx$ 22 m$\Omega$, giving a very respectable MR of $(R_{AP} - R_P)/R_P \approx$ 14%. In contrast to the V notch of Fig. 2 where one trapping step occurs in the thin Py layer, here we see several intermediate trapping steps that commence once the wall is injected into the wire at $|H| \sim$ 10 Oe and rare trapping of the wall at the notch in the thick Py layer. Also the direction of field sweep affects these steps. As opposed to the V-notch case where the wall-pinning site is well defined, the more elongated U-notch (see Fig.1 (c)-(i)) can have multiple pinning sites associated with wire-edge roughness and other defects. After repeated field sweeps from –300 Oe to +300 Oe, $\Delta R$ became more stable. The open circles represent one of these more stable sweeps that happens to exhibit wall trapping in the thick Py layer. The letters B through G indicate the field at which a positive field sweep was interrupted and then R was monitored for long periods of time to assess the stability of wall trapping. Then the field sweep was started again at –300 Oe and stopped at another value of H.

For the sample in Fig. 4, the CPP current is less uniform because the Au top contact is normal. The multiple trapping steps shown in Fig. 4 are very similar to what is seen at 4.2 K in U-notch samples with Nb top contacts where the CPP current is more uniform. Note also that, in Fig. 3 at 295K, the CPP current was not uniform. Thus the uniformity of the CPP current seems not to be an important parameter for detecting domain wall motion.

Fig. 5 shows the time dependence of R after the field sweep was stopped at the values of H indicated on Fig. 4 as B – G and in Fig. 2 as A. For the latter, the stability of the trapped wall in the V-notch of the thin Py layer was of interest. Clearly curve A shows the best stability, as expected. Curves B – G exhibit significant changes in R with time, reflecting the weaker pinning of the walls in the

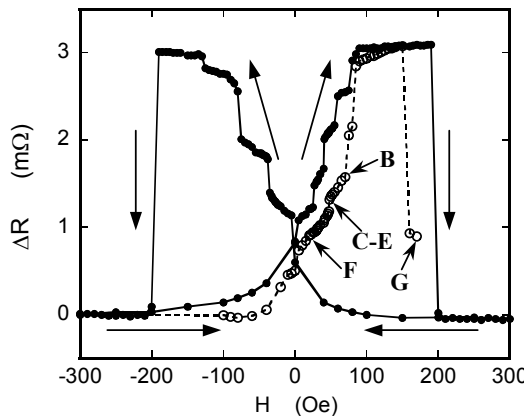

Fig. 4. $\Delta R (= R - R_P)$ vs H at 4.2 K for a sample with a U-shaped notch (solid circles) and a Au top contact. Open circles: H sweeps that start at –300 Oe and then stop for over 1000 s at a given H represented by B – G. Then the field sweep is started again at –300 Oe.

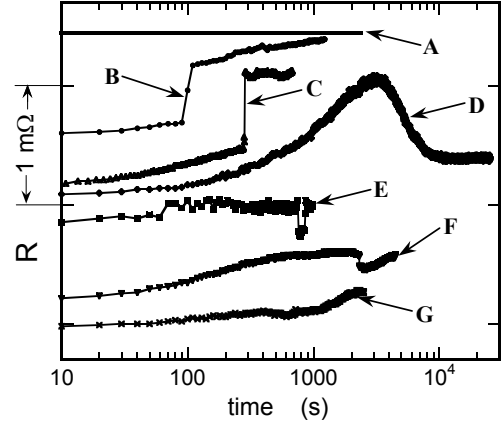

Fig. 5. Resistance R vs time, commencing 10 s after magnetic field sweep is stopped. In Fig. 2, A shows this value of H; and likewise, in Fig. 4, B – G show these values of H. The ordinate values of these curves have been shifted somewhat for clarity, but they share the same incremental resistance scale.

U-notches. Studies of standard resistors at 4.2 K indicated that long-term drifts of the electronics were $\leq$ 0.02 m$\Omega$, significantly less than these changes seen in Fig. 5. These domain wall motions in the U-notches are somewhat similar to what was seen in very narrow Ni wires where tunneling through a barrier below 2 K was proposed,[9] as opposed to thermal activation over a barrier. Since our data were only taken at 4.2 K, we presume the latter mechanism takes place.

In conclusion, we have used localized perpendicular-current giant magnetoresistance to detect the occurrence and motion of a trapped domain wall in a narrow ferromagnetic wire. This versatile detection technique will allow a wide range of interesting wall-trapping geometries to be investigated.

We acknowledge preliminary work by R. Slater, useful advice by J. Bass and S. Urahzdin, and support by the Keck Microfabrication Facility and NSF grant DMR-02-02476.

[1] T. Ono, H. Miyajima, K. Shigeto,and T Shinjo, Appl. Phys. Lett. **72**, 1116 (1998).
[2] K. Shigeto, T. Okuno, T. Shinjo, Y. Suzuki, and T. Ono, J. Appl. Phys. **88**, 6636 (2000) and references therein.
[3] J. Grollier, P. Boulenc, V. Cros, A. Hamzić, A. Vaurès, A Fert and G. Faini, Appl. Phys. Lett. **83**, 509 (2003).
[4] M. Kläui, C. A. F. Vaz, J. A. C. Bland, W. Wernsdorfer, G. Faini, E. Cambril and L. J. Heyderman, Appl. Phys. Lett. **83**, 105 (2003).
[5] M. Tsoi, R. E. Fontana and S. S. P. Parkin, Appl. Phys. Lett. **83**, 2617 (2003).
[6] J. Bass and W. P. Pratt, Jr., J. of Magn. Magn. Mater. **200**, 274 (1999).
[7] R. D. McMichael, J. Eicke, M.J. Donahue and D. G. Porter, J. Appl. Phys. **87**, 7058 (2000).
[8] R. D. McMichael, and M.J. Donahue, IEEE Trans. Magn. **33**, 4167 (1997).
[9] K. Hong and N. Giordano, J. Magn. Magn. Mater. **151**, 396 (1995) and references therein.